\documentclass[superscriptaddress,amsmath, amssymb, amsfonts, twocolumn, footinbib]{revtex4-1}
\usepackage[german,american,english]{babel}
\usepackage{graphicx}
\usepackage{graphics}
\usepackage{dcolumn}
\usepackage{bm}
\usepackage{amssymb}
\usepackage{amsmath}
\usepackage{amsfonts}
\usepackage{epsfig}
\usepackage{mathbbol}
\usepackage{latexsym}
\usepackage{dcolumn}
\usepackage{subfigure}
\usepackage{hyperref}
\usepackage{float}
\usepackage[normalem]{ulem}
\usepackage[usenames, dvipsnames]{color}
\usepackage{epstopdf}
\usepackage{comment}
\hypersetup{                    
    colorlinks,
    citecolor=red,
    filecolor=red,
    linkcolor=red,
    urlcolor=red
}

 \newcommand {\beq}{\begin{equation}}
\newcommand {\eeq}{\end{equation}}
\newcommand {\beqn}{\begin{eqnarray}}
\newcommand {\eeqn}{\end{eqnarray}}
\newcommand {\bit}{\begin{itemize}}
\newcommand {\eit}{\end{itemize}}

\newcommand{\ba}{\begin{array}{rl}}
\newcommand{\ea}{\end{array}}
\newcommand{\bc}{\begin{cases}}
\newcommand{\ec}{\end{cases}}

\newcommand{\om}{\iffalse}

\definecolor{mygray}{gray}{0.6}
\definecolor{gold}{RGB}{150, 150, 10}
\definecolor{mygreen}{RGB}{40, 200, 100}

\begin{document}
\renewcommand{\vec}{\mathbf}
\renewcommand{\Re}{\mathop{\mathrm{Re}}\nolimits}
\renewcommand{\Im}{\mathop{\mathrm{Im}}\nolimits}
\title{Topological spin-plasma waves}

\author{Dmitry K. Efimkin}
\email{dmitry.efimkin@monash.edu}
\affiliation{School of Physics and Astronomy, Monash University, Victoria 3800, Australia}
\affiliation{ARC Centre of Excellence in Future Low-Energy Electronics Technologies, Monash University, Victoria 3800, Australia}
\author{Mehdi Kargarian}
\email{kargarian@physics.sharif.edu}
\affiliation{Department of Physics, Sharif University of Technology, Tehran 14588-89694, Iran}

\begin{abstract}
The surface of a topological insulator hosts Dirac electronic states with the spin-momentum locking, which constrains spin orientation perpendicular to electron momentum. As a result, collective plasma excitations in the interacting Dirac liquid manifest themselves as coupled charge- and spin-waves. Here we demonstrate that the presence of the spin component enables effective coupling between plasma waves and spin waves at interfaces between the surface of a topological insulator and insulating magnet. Moreover, the helical nature of spin-momentum locking textures provides the phase winding in the coupling between the spin and plasma waves that makes the spectrum of hybridized spin-plasma modes to be topologically nontrivial. We also show that such topological modes lead to a large thermal Hall response. \end{abstract}
\date{\today}
\maketitle

\section{Introduction}
The search for new materials and experimentally realizable heterostructures harboring 
topological quantum phases of matter has become a central paradigm in condensed matter physics in past 
few decades. Some examples include, but not restricted to, discovery of topological insulators in 3D bulk materials \cite{Hasan:RMP2010, Hasan:Annals2011} and in 2D quantum wells \cite{Bernevig1757, Konig:Science2007}, realization 
of Majorana bound states in topological superconducting heterostructures \cite{Fu:PRL2008, Lutchyn:PRL2010, Oreg:PRL2010, Sau:PRB2010, Alicea_2012, Beenakker:2013} as a promising platform for topological quantum computations \cite{DasSarma:RMP2008}, topological Mott insulators \cite{Pesin:Nat2010}, topological crystalline insulators \cite{Hsieh:ncomm2012, Kargarian:PRL2013}, and topological Weyl and Dirac semimetals \cite{Burkov:NatMat2016, Armitage:RMP2018, Burkov:Annals2018}. The appearance of topologically protected gapless surface and edge states is a direct consequence of topological electron states in the bulk. In another frontier, the notion of bulk band topology has 
been extended to include non-electron systems such as photonic systems \cite{Raghu:PRA2008, Haldane:PRL2008, Wang:nat2009, Hafezi:NatPhys2011, Khanikaev:NatMat2013}, polaritons \cite{TopPolaritons1,TopPolaritons2}, phonons \cite{Prodan:PRL2009, Zhang:PRL2010, Yang:PRL2015, Susstrunk:pnas2016, Liu:PRB2017}, magnons \cite{Katsura:PRL2010, Matsumoto:PRL2011, Matsumoto:PRB2011, Shindou:PRB2013, Shindou2:PRB2013, Owerre_2016, Thingstad:PRL2019}, magnetoelastics \cite{Takahashi:PRL2016, MagnonPhonon1,MagnonPhonon2,MagnonPhonon3,MagnonPhonon5,MagnonPhonon6,MagnonPhonon7},  and recently plasmons \cite{MagnetoPlasmons1,MagnetoPlasmons2}. In all of these systems, which are described by bosonic collective modes, the band topology emanates from the nontrivial Berry curvature  
of the underlying Bloch wave description of bulk modes, which upon integration over the momentum space 
leads to an integer topological index. 

The hybridization between different bosonic collective modes may lead to new physical phenomena with intriguing applications in constructing electronic, optical, and thermal devices. One example is the coupling between magnons and phonons, the formation of magnon polarons, due to spin-lattice interactions at low temperatures \cite{Kittel:PMP1949, Kittel:PR1958}. This coupling inspires the use of sound-induced magnetization dynamics \cite{Kamra:PRB2015} and acoustic spin pumping in designing the spin \cite{Uchida:NatMat2011, Weiler:PRL2012} and energy transport devices \cite{Kikkawa:PRL2016, Flebus:PRB2017}, and the electric field control of spin  currents in multiferroic magnonics \cite{Sigrist:PRL2015}. Also, it is shown that in magnets with easy-axis anisotropy and strong  Dzyaloshinskii-Moriya interaction the coupling between magnons and phonons induce thermal Hall effects with possible applications in spin caloritronics  \cite{Bauer:NatMat2012}.

\begin{figure}[t]
	\vspace{-0.1in}
	\begin{center}
		\includegraphics[width=8.0cm]{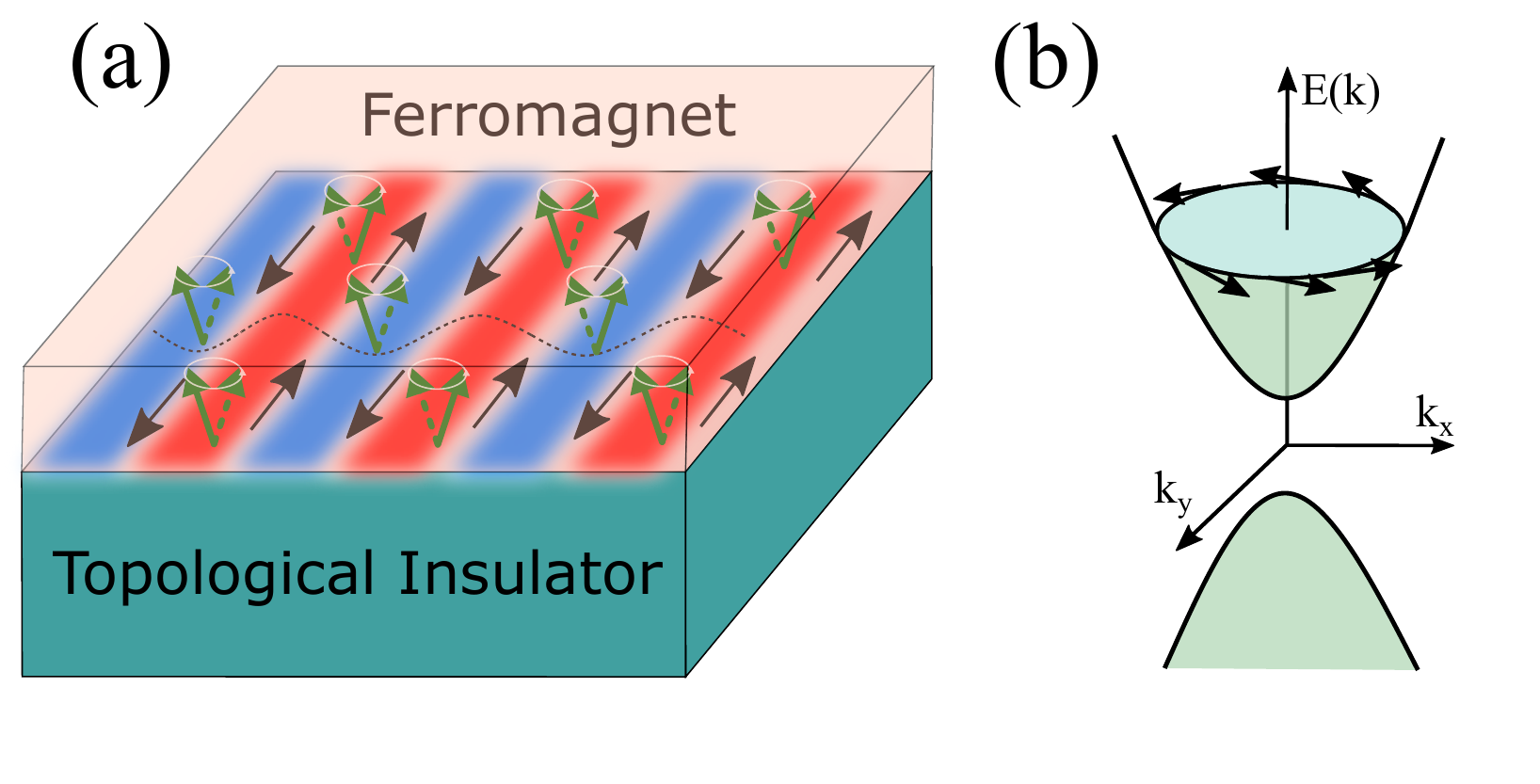}
		\caption{\label{TIMag} (a) shows an interface between the topological insulator and ferromagnetic thin film. Due to the spin-momentum locking for Dirac electrons that is illustrated in (b) plasma waves manifest themselves as coupled density (red and blue denote regions with excess and deficit of electrons) and transverse spin (black arrows) waves. The latter enables the effective coupling with fluctuating magnetic moments (vertical green arrows) and formation of the hybrid spin-plasma waves.} 
	\end{center}
	\vspace{-0.1in}
\end{figure}

In this letter we introduce a novel hetrostructure that is sketched in Fig.~\ref{TIMag} and consists of a topological insulator and an insulating magnet. Due to the helical nature of Dirac electron liquid, plasma waves hosted by it are accompanied by the transverse spin-wave~\cite{SpinPlasmonsRaghu}.
We show that when these topologically featureless modes are coupled to each other, the hybrid system is topologically nontrivial. 
Our model is different from chiral Berry plasmons \cite{Song:pnas2016}, where the boundary modes 
are not topological modes and arise due to the split in energy dispertions of oppositely directed plasmon waves. And in contrast to the topological magnetoplasmon \cite{MagnetoPlasmons1}, our model doesn't require a magnetic field which is rather impeding 
in devices. Moreover, we show that these hybrid topological modes give rise to a large thermal Hall response which can be measured experimentally. Our findings open a new experimental and theoretical avenue to explore the topological phases of matter even in 
trivial bosonic and classical systems when combined appropriately.



\section{Model}
 Consider an interface between a magnetic thin film and the surface of a topological insulator (TI) as shown in Fig.~\ref{TIMag}(a). We assume that the magnet is insulator and has an easy-axis anisotropy. The latter dictates magnetic moments to be ordered perpendicular to the TI surface, e.g. in the $\vec{e}_\mathrm{z}$-direction. On the other hand, propagating magnetic fluctuations have only in-plane component $\vec{l}_{t\vec{r}}=l^\mathrm{x}_{t\vec{r}} \vec{e}_\mathrm{x}+ l^\mathrm{y}_{t\vec{r}} \vec{e}_\mathrm{y}$ and are known as spin-waves or magnons. They interact with interacting Dirac electron liquid at the TI surface that is known to host collective plasma excitations or plasmons \cite{SpinPlasmonsRaghu}. The classical picture of spin and plasma waves is physical and more intuitive and will be used in here, while the quantum description is presented in Appendices A and B. Importantly, spin and plasma waves do not couple directly but only through the degenerate quantum Dirac helical liquid that is described by the following Hamiltonian 

\begin{equation}
H=v [\vec{p}\times \mathbf{\sigma}]_z + \Delta \sigma_\mathrm{z}- \epsilon_\mathrm{F} + \Delta \mathbf{\sigma} \cdot \vec{l}_{t\vec{r}} + e \phi_{t\vec{r}},  
\end{equation}
where $v$ and $\epsilon_\mathrm{F}$ are velocity and Fermi energy of Dirac electrons. The Hamiltonian acts at the spinor wave function $\psi_{t\vec{r}}=\{\psi^\uparrow_{t\vec{r}},\psi^\downarrow_{t\vec{r}}\}^T$ for electrons. The energy $\Delta$ determines their coupling strength with magnetic moments. 
The time-dependent scalar potential $\phi_{t\vec{r}}$ is created by  the density fluctuations of Dirac electron liquid that accompany plasma-waves. 

The dynamics of magnetic fluctuations $\vec{l}_{t\vec{r}}$ follow the linearized Landau-Lifshitz-Gilbert equation \cite{LL} (See also Appendix B for its derivation) given by 
\begin{equation}
\label{LL1} 
\rho_\mathrm{s} \left[\partial_t \vec{l}_{t\vec{r}} \times \vec{e}_\mathrm{z}\right]= \rho_\mathrm{s}\epsilon_{\hat{\vec{p}}} \vec{l}_{t \vec{r}} + \Delta \vec{s}_{t\vec{r}}, 
\end{equation}
where $\epsilon_\vec{p}=\delta_\mathrm{s}+\vec{p}^2/2 m_s$ is the dispersion of spin-waves with mass $m_\mathrm{s}$ and the gap $\delta_\mathrm{s}$ induced by anisothropy. $\rho_\mathrm{s}$ is the density of magnetic moments in the magnet. They are coupled with the spin density $\vec{s}_{t\vec{r}}=s^\mathrm{x}_{t\vec{r}} \vec{e}_\mathrm{x}+ s^\mathrm{y}_{t\vec{r}} \vec{e}_\mathrm{y}$ of Dirac liquid and can be excited by its oscillations. 

The scalar potential is determined self-consistently by electron density $\rho_{t\vec{r}}$ and satisfies the Poisson equation $\Delta \phi_{t\vec{r}}=-4\pi e \rho_{t\vec{r}}$. Its solution can be presented~\footnote{We neglect photon retardation effects that are known to promote plasma waves, or plasmons, to plasmon-polaritons. It does not modify the results and becomes important only at very low momenta $c q\lesssim \omega_\vec{q}$ that is much smaller then avoided crossing of dispersion curves for spin and plasma waves} in a compact way as      
\begin{equation}
\label{Poisson}
e \phi_\vec{t \vec{r}}=\int d\vec{r}' V_{\vec{r}-\vec{r}'} \rho_\vec{t \vec{r}'}.
\end{equation}
The potential $V_\vec{r}$ incorporates details of a dielectric screening and the sample geometry. 
For the sake of simplicity we use $V_\vec{r}= e^2/\kappa r$ with $\kappa$ is the effective dielectric constant of the interface.  


The dynamics of $l^x_{t \vec{r}}$ and $l^y_{t \vec{r}}$ in Eq.~(\ref{LL1}) is mutually coupled and it is instructive to introduce complex fields as $l^-_{t\vec{r}}=l^\mathrm{x}_{t\vec{r}}-i l^\mathrm{y}_{t\vec{r}}$ and $l^+_{t\vec{r}}=l^\mathrm{x}_{t\vec{r}}+i l^\mathrm{y}_{t\vec{r}}$. Their dynamics is governed by 
\begin{equation}
	\label{LL3}
\rho_\mathrm{s} (\mp i \partial_t l^\pm_{t \vec{r}}-\epsilon_{\hat{\vec{p}}}l^\pm_{t \vec{r}})=\Delta s^\pm_{t \vec{r}}
\end{equation}
with  $s^\pm_{t \vec{r}}=(s^x_{t \vec{r}} \pm i s^y_{t \vec{r}})/2$.  After Fourier transform, Eqs.~\eqref{Poisson} and \eqref{LL3} can be presented in a compact matrix form as
\begin{equation*}
\hat{L}_{\omega \vec{q}}^0 f_{\omega \vec{q}}=m_{\omega \vec{q}}, \; \; L_{\omega \vec{q}}^0=
\begin{pmatrix}
\frac{1}{V_\vec{q}} & 0 & 0 \\
0 & \frac{\rho_s(\omega-\epsilon_\vec{q})}{\Delta^2}   & 0 \\
0 & 0 & -\frac{\rho_s (\omega+\epsilon_\vec{q})}{\Delta^2}  \\
\end{pmatrix}.
\end{equation*}
Here we have introduced the vectors for \emph{fields} $f_{\omega \vec{q}}=\{e \phi_{\omega \vec{q}}, \Delta l^-_{\omega \vec{q}}, \Delta l_{\omega ,\vec{q}}^+\}$ and \emph{the matter densities} $m_{\omega \vec{q}}=\{\rho_{\omega \vec{q}},s_{\omega \vec{q}}^-, s_{\omega \vec{q}}^+\}$ with  $s^\pm_{\omega \vec{q}}=(s^x_{\omega \vec{q}} \pm i s^y_{\omega \vec{q}})/2$. The matrix $\hat{L}^0_\vec{\omega q}$ can be interpreted as the inverse Green function that describes response of fields $f_{\omega\vec{q}}$ to matter oscillations $m_{\omega\vec{q}}$. A closed form of equations can be derived by closely following the ideas of the dynamical mean field theory~\cite{Mahan,BruusFlensberg}. The main idea is that the fields $\hat{f}_{\omega \vec{q}}$ are not only produced by the matter, but also influence it in the self-consistent manner. The response of the matter $m_{\omega \vec{q}}$ to the fields $f_{\omega \vec{q}}$ can be presented as follows 
\begin{equation}
m_{\omega \vec{q}}=\hat{\Pi} f_{\omega \vec{q}}, \quad \quad  
\hat{\Pi}=
\begin{pmatrix}
\Pi^{00}_{\omega \vec{q}} & \Pi^{0+}_{\omega \vec{q}} & \Pi^{0-}_{\omega \vec{q}} \\
\Pi^{-0}_{\omega \vec{q}} & \Pi^{-+}_{\omega \vec{q}} & \Pi^{--}_{\omega \vec{q}} \\
\Pi^{+0}_{\omega \vec{q}} & \Pi^{++}_{\omega \vec{q}} & \Pi^{+-}_{\omega \vec{q}} \\
\end{pmatrix},
\end{equation}
where the entities are the density-density $\Pi^{00}_{\omega \vec{q}}$, spin-spin $\Pi^{\pm \pm}_{\omega \vec{q}}$ and $\Pi^{\mp \pm}_{\omega \vec{q}}$, and the cross-correlated $\Pi^{\pm 0}_{\omega \vec{q}}$ spin-density response functions. The latter ones 
provide the coupling between spin and plasma waves. We eliminate the matter $\hat{m}_\vec{q}$ and obtain a closed system of equations for the fields $\hat{f}_{\omega \vec{q}}$ as $(\hat{L}^0_{\omega \vec{q}}-\hat{\Pi}_{\omega \vec{q}})f_{\omega \vec{q}}=0$. It has nontrivial solutions only if its determinant, the dispersion equation, vanishes:   
\begin{equation}
\label{DispersionEq}
\mathrm{det}[\hat{L}^0_{\omega \vec{q}}-\hat{\Pi}_{\omega \vec{q}}]=0,
\end{equation}
which determines the dispersion of the hybrid spin-plasma waves. 


\section{Spin-density response function}
The coupling between spin- and plasma- waves is determined by the spin-density response functions $\Pi_{\omega \vec{q}}^{\pm0}$. 
For conventional electrons with quadratic dispersion, $\Pi_{\omega \vec{q}}^{\pm 0}$ =0, making spin and plasma waves to be decoupled; the plasma waves in this case manifests themselves as \emph{purely charge density oscillations} that are not coupled with spin-waves. 

This is not the case for helical Dirac electrons at the surface of TI. The spin-momentum locking results in the relation 
$\vec{s}_{t\vec{r}}=[\vec{j}_{t\vec{r}}\times \vec{e}_z]/v$ between particle current $\vec{j}_{t\vec{r}}$ and spin density $\vec{s}_{t\vec{r}}$. As a result, plasma waves at the surface of a TI were shown to  manifest themselves as coupled longitudinal charge-density and transverse spin-density waves~\cite{SpinPlasmonsRaghu,SpinPlasmonsEfimkin}, as it is sketched in Fig.~\ref{TIMag}(a). In our magnetic heterostructure the spin component of plasma couple to the spin-waves of the ferromagnetic layer. Moreover, the outlined above relation accompanied by the particle conservation law $\partial_t\rho+\mathrm{div}\vec{j}=0$ establishes the \emph{exact} relation between $\Pi_{\omega \vec{q}}^{\pm0}$ and $\Pi_{\omega \vec{q}}^{00}$. The fluctuations of particle density $\rho_{\omega \vec{q}}$ generate the longitudinal current $\vec{j}_{\omega \vec{q}}=e \omega \vec{n}_{\vec{q}}\rho_{\omega \vec{q}}/q$ with $n_\vec{q}=\vec{q}/q$ and therefore the transverse spin density $\vec{s}_{\omega \vec{q}}$ reads as follows 
\begin{equation}
\vec{s}_{\omega \vec{q}}=  [\vec{e}_\mathrm{z} \times \vec{n}_\vec{q} ] \frac{\omega}{vq} \rho_{\omega \vec{q}}.    
\end{equation}
If we reintroduce $s^\pm_{\omega \vec{q}}=(s^x_{\omega \vec{q}}\pm i s^y_{\omega \vec{q}})/2$, its connections with $\rho_{\omega \vec{q}}$ dictates the following identity    
\begin{equation}
\label{PiRelation}
\Pi_{\omega \vec{q}}^{\pm 0}=\pm\frac{ie^{\pm i\phi_\vec{q}}}{2}\frac{\omega}{vq} \Pi_{\omega \vec{q}}^{00},    
\end{equation}
The critical observation is that spin-density response function $\Pi_{\omega \vec{q}}^{\pm 0}$ has the phase winding factor. This cornerstone relation of our theory ensures the nontrivial topology of hybridized spin-plasma waves.

\section{Dispersion of spin-plasma waves}
To proceed further, we assume that the Dirac liquid is degenerate, $T\ll\epsilon_\mathrm{F}$, and focus at the long-wave, $q\ll p_\mathrm{F}$, and the low-frequency regime $\omega\ll\epsilon_\mathrm{F}$. The calculation of cross-correlated response functions $\hat{\Pi}_{\omega \vec{q}}$ within the random phase approximation (RPA) is presented in Appendix C. In particular, the RPA respects the general relation \eqref{PiRelation} and the density-density response function is given by 
\begin{equation}
\Pi_{\omega\vec{q}}^{00}=N_\mathrm{F}\left(\frac{\omega}{\sqrt{(\omega+i \delta)^2-u^2 \vec{q}^2}}-1\right).
\end{equation}
Here $N_\mathrm{F}=p_\mathrm{F}/2\pi u \hbar$ is the density of states at the Fermi level and $u=h v$ is the Fermi velocity with the factor $h=vp_\mathrm{F}/\epsilon_\mathrm{F}$ that reflects the presence of small gap $\Delta\ll\epsilon_\mathrm{F}$ in the Dirac spectrum. In the absence of coupling with spin-waves, the dispersion relation reduces to $1-V_\vec{q} \Pi_{\omega\vec{q}}^{00}=0$ and gives the dispersion relation for plasma waves 
\begin{equation}
\label{DispersionPlasma}
\omega_\vec{q}^2=u^2 \vec{q}^2 \frac{(N_\mathrm{F} V_\vec{q}+1)^2}{2 N_\mathrm{F} V_\vec{q}+1}.
\end{equation}
that is presented schematically in Fig.~\ref{TIMag}(a). At small momenta it has the square root behavior, $\omega_\vec{q}=\sqrt{u^2 \vec{q}^2 N_\mathrm{F} V_{\vec{q}}/2}\propto \sqrt{q}$, well known for two-dimensional electrons (conventional or Dirac). At larger momentum $q$, it approaches the continuum of electron-hole excitations of the Dirac electronic liquid $\omega<uq$ that reflects itself in non-zero imaginary part of $\Pi_{\omega\vec{q}}^{00}$ and provides the Landau damping to any modes that enter into it.

\begin{figure}[t]
	\label{FigDispersions}
	\vspace{-0.0in}
	\begin{center}
		\includegraphics[width=8.0cm]{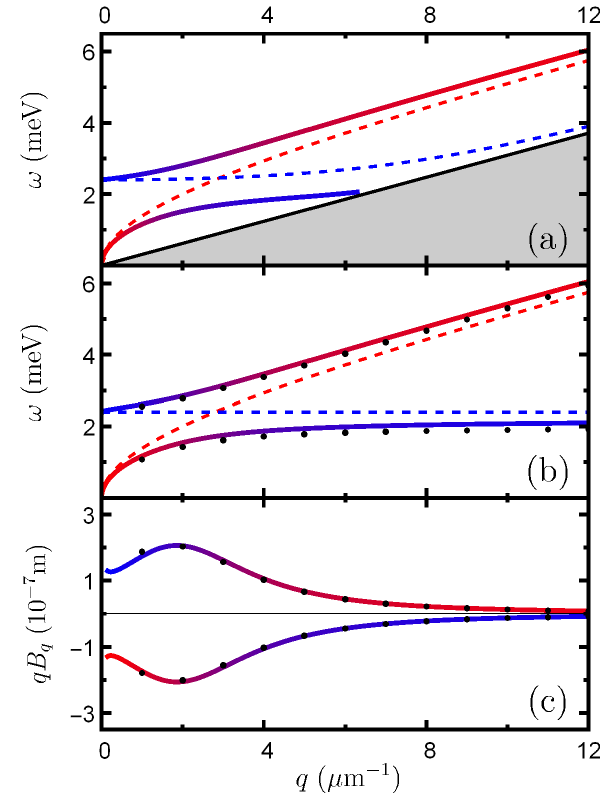}
		\vspace{-0.1in}
		\caption{\label{PlotResults} The colored lines in (a) and (b) present the dispersion curves of hybrid spin-plasma waves calculated using (a) the dispersion relation, Eq.~\eqref{DispersionEq} and (b) the truncated Hamiltonian $H_\vec{q}$, Eq~\eqref{BdG1}. 
		The dashed curves are the dispersion of bare spin and plasma waves neglecting the coupling between modes. (c) Berry curvature of the hybrid spin-plasma waves calculated using $H_q$. The black dots in (b) and (c) correspond to the predictions based on the full BdG Hamiltonian $K_{\vec{q}}$.} 
	\end{center}
	\vspace{-0.3in}
\end{figure}

If we approximate the dispersion of spin-waves to be flat $\epsilon_\vec{q}=\delta_\mathrm{s}$, the dispersion of the hybrid spin-plasma waves, that satisfy Eq.~(\ref{DispersionEq}), depends on three controlling parameters: 1) the modified fine structure constant for Dirac electrons $\alpha=h^2e^2/\hbar v  \kappa$; 2) the ration between the coupling energy at the interface $\epsilon_\mathrm{\Delta}=N_\mathrm{F}h^2 \Delta^2/16 \rho_\mathrm{s}$ and the Fermi one $g^2=\epsilon_\mathrm{\Delta}/\epsilon_\mathrm{F}$; 3) the dimensionless gap in the spectrum of spin-waves $d=\delta_\mathrm{s}/\epsilon_\mathrm{F}$. We employ the following set of parameters $\epsilon_\mathrm{F}\approx 120 \; \hbox{meV}$, $v\approx0.5 \; 10^6\; \hbox{m}/\hbox{s}$, $\kappa\approx80$, $\rho_\mathrm{S}\approx 2\;10^{12}\; \hbox{cm}^{-2}$, $\Delta\approx 20~\hbox{meV}$, and $\delta_\mathrm{s}\approx 2.4~\hbox{meV}$ that are relevant for recently discovered magnetic TI $\hbox{MnBi}_2\hbox{Te}_4$~\cite{MagneticTI1,MagneticTI2,MagneticTI3,MagneticTI3}. The resulting controlling parameters are $\alpha\approx0.1$, $g\approx 0.04$ and $d\approx 0.02$. The smallness of $\alpha$ and $g$ ensures the applicability of the RPA. Importantly, the contribution of $\Pi_{\omega \vec{q}}^{\pm\pm}$ and $\Pi_{\omega \vec{q}}^{\mp\pm}$ that result in renormalization of the bare spin-waves by interactions with Dirac liquid~\cite{TIMagnetism1,TIMagnetism2,TIMagnetism3,TIMagnetism4,TIMagnetism5,TIMagnetism6,TIMagnetism7} are of the second order in $g$. It is much smaller then the contribution of $\Pi^{\pm0}_{\omega \vec{q}}$ that is of the first order in $g$ and is responsible for the coupling between spin and plasma waves. 

The Fig.~\ref{PlotResults}(a) presents the dispersion curves for hybrid spin-plasma waves accompanied by their \emph{bare} counterparts (calculated assuming $\Pi_{\omega \vec{q}}^{\pm0}=0$). The curve for the bare plasma wave follows Eq.~\eqref{DispersionPlasma}. The one for bare spin-wave is almost dispersionless, $\epsilon_\vec{q}\approx\delta_\mathrm{s}$ and is bended by the interactions with Dirac liquid only in the vicinity of the continuum of electron-hole excitations $q_\mathrm{s}\approx \delta_\mathrm{s}/v$. 
The Fig.~\ref{PlotResults}(a) clearly demonstrates the effective coupling between spin- and plasma- waves provided by spin-density response function $\Pi^{\pm0}_{\omega \vec{q}}$. Their hybridization is especially effective in the vicinity of the avoiding crossing $q_*\approx  2 d^2 p_\mathrm{F}/\alpha$. The nontrivial topology of hybrid spin-plasma waves is encoded in the corresponding eigenstates of the dispersion equation are nor apparent yet.

\section{Nontrivial topology of spin-plasma waves}
 To uncover the nontrivial topology of hybrid spin-plasma waves, two simplifications are in order: 1) The plasma-pole approximation, $1-V_\vec{q}\Pi_{\omega \vec{q}}^{00}\approx 1-\omega_\vec{q}^2/\omega^2$, where the plasma frequency $\omega_\vec{q}$ is given by Eq.~\eqref{DispersionPlasma}. The approximation is known to work very well outside the continuum and becomes exact at $\omega\gg uq$. 2) We set $\Pi^{\pm\pm}_{\omega \vec{q}}=\Pi^{\pm\mp}_{\omega \vec{q}}=0$. Their effects on the bare spin-waves is of the second order in the small parameter $g$ and they become important only in the vicinity of the continuum. 

Using these simplifications and the transformation,
$l^\pm_{t\vec{q}}=a^\pm_{t\vec{q}}/\sqrt{\rho_\mathrm{s}}$, and $e\phi_{t \vec{q}}=-\sqrt{V_\vec{q}} \dot{\phi}_{t\vec{q}}'$, the classical equations, Eq.~\eqref{DispersionEq}, can be written as 
\begin{equation}
\label{OscillatorSystem}
\begin{split}
    &\omega^2 \phi_{\omega\vec{q}}'=\omega_{\vec{q}}^2 \phi_{\omega\vec{q}}'+ \sqrt{2 \omega_{\vec{q}}}  M_{\vec{q}} (a^-_{\omega \vec{q}} + a^+_{\omega,\vec{q}}), \\
    &\omega a^{\mp}_{\omega\vec{q}} = \pm\sqrt{2 \omega_{\vec{q}}} M_{\vec{q}}^* \phi_{\omega\vec{q}}'\pm\epsilon_\vec{q}a^{\mp}_{\omega\vec{q}}. 
\end{split}
\end{equation}
Here $M_\vec{q}=\sqrt{\omega_\vec{q} \epsilon_\mathrm{\Delta}}e^{i \phi_\vec{q}}$ is the matrix-element of the coupling between spin and plasma- waves and inherits the phase winding factor from the spin-density response function. In the time domain, the equation for $\phi'_{t\vec{q}}$ represent the harmonic oscillator with an external force induced by spin-waves and frequency $\omega_\vec{q}$.  However, the equation for $a^-_{\omega\vec{q}}$ and $a^+_{\omega,\vec{q}}$ are of the first order and is a classical analogue of the quantum Schrodinger wave equation~\footnote{It should be noted that they are not independent because $a^+_{t \vec{r}}=(a^-_{t \vec{r}})^*$ ensures $a^+_{\omega \vec{q}}=(a^-_{-\omega, - \vec{q}})^*$ }. This important observation bridges us towards the topological analysis of spin-plasma waves. Doing so, we have to rewrite the harmonic oscillator as two coupled equations of the first order. The naive way of using, $\partial_t p_{t\vec{q}}=-\omega_\vec{q}^2 \phi'_{t\vec{q}}+F_{t\vec{q}}$ and $\partial_t\phi'_{t\vec{q}}=p_{t\vec{q}}$, does not has the Schrodinger structure. We introduce a complex combination of $\partial_t\phi'_{t\vec{q}}$ and $\phi'_{t\vec{q}}$ as  
\begin{equation}
\begin{split}
&b^-_{t\vec{q}}=\frac{1}{\sqrt{2 \omega_\vec{q}}} \left[\omega_\vec{q}\phi'_{t\vec{q}}+ i \partial_t \phi'_{t\vec{q}} \right], \\  
&b^+_{t,-\vec{q}}=\frac{1}{\sqrt{2 \omega_\vec{q}}} \left[\omega_\vec{q}\phi'_{t,-\vec{q}}- i \partial_t \phi'_{t,-\vec{q}} \right]. 
\end{split}
\end{equation}
We combine degrees of freedom for spin and plasma waves as  $\psi_{\omega\vec{q}}=\{\psi^-_{\omega\vec{q}},\psi^+_{\omega\vec{q}}\}$ with $\psi^-_{\omega\vec{q}}=\{b^-_{\omega\vec{q}}, a^-_{\omega\vec{q}}\}$ and $\psi^+_{\omega\vec{q}}=\{b^+_{\omega,\vec{q}}, a^+_{\omega,\vec{q}}\}$. The system of Eqs.~\eqref{OscillatorSystem} can be presented as Schrodinger-like equation with the bosonic Bogoliubov-de Gennes (BdG) dynamical matrix~\footnote{The dynamical matrix is connected with the BdG Hamiltonian $H_\vec{q}^\mathrm{BdG}$ as follows $\hat{H}_\vec{q}^\mathrm{BdG}=\hat{\Sigma}_\mathrm{z} \hat{K}_\vec{q}$ with $\Sigma_\mathrm{z}=\mathrm{diag}[1,1,-1,-1]$ is the generalized Pauli matrix. $H_\vec{q}^\mathrm{BdG}$ can be really interpreted as a Hamiltonian.  If we quantize the waves, $a^-_{t\vec{q}}$ and $b^-_{t\vec{q}}$ will be promoted to annihilation operators of magnons and plasmons, that are quantum counterpars of spin and plasma waves. The equation, (\ref{BdG1}) being rewritten in the time domain represents the time-dependent Heisenberg equation with $\hat{H}_\vec{q}^\mathrm{BdG}$ is its quantum Hamiltonian.} as follows 
\begin{equation}
\label{BdG1}
K_\vec{q} \psi_{\omega \vec{q}} = \omega \psi_{\omega \vec{q}} , \quad\quad  
K_\vec{q}=\begin{pmatrix}
H_{\vec{q}}  & Z_{\vec{q}} \\ -Z_{\vec{q}}^\dagger & -H_{-\vec{q}}^*
\end{pmatrix}.
\end{equation}

It is non-Hermitain, since it is a paraunitary transformed bosonic Hamiltonian, and its blocks are given by 
\begin{equation}
\label{BdG2}
H_\vec{q}=\begin{pmatrix}
\omega_\vec{q} & M_\vec{q}  \\
M_\vec{q}^* & \epsilon_\vec{q}
\end{pmatrix}, \quad \quad 
Z_\vec{q}=\begin{pmatrix}
0 & M_{-\vec{q}}^*  \\
M_{\vec{q}}^* & 0
\end{pmatrix}.
\end{equation}
In the absence of the coupling, $M_\vec{q}=0$, the BdG dynamical matrix is diagonal $K_{\vec{q}}=\mathrm{diag}[\omega_\vec{q},\epsilon_\vec{q},-\omega_\vec{q},-\epsilon_\vec{q}]$ and describes bare spin- and plasma waves supplemented by spurious negative energy branches that are not dynamically independent. The diagonal blocks in $K_\vec{q}$ describe the resonant coupling between branches with energies of the same sign, while the term $Z_{\vec{q}}$ corresponds to the off resonant coupling between positive and energy ones.   

The spectrum of positive energy states for $K_\vec{q}$ is given by 
\begin{equation}
	\label{SPDispersion}
	\omega_{\pm}^2=\frac{\omega_\vec{q}^2+\epsilon_\vec{q}^2}{2}\pm \sqrt{ \left(\frac{\omega_\vec{q}^2-\epsilon_\vec{q}^2}{2}\right)^2 + 4 \epsilon_\vec{q} \omega_\vec{q} |M_\vec{q}|^2}, 
\end{equation}
They are plotted in Fig.~\ref{PlotResults}(b) and well approximate the curves in Fig.~\ref{PlotResults}(a) that has been calculated using the dispersion equation, Eq.~\eqref{DispersionEq}, except in the vicinity and within the particle-hole continuum of the electron liquid. The coupling between spin and plasma waves pushes their lower hybrid mode towards $\omega=0$. According to Eq.~\eqref{SPDispersion}, the touching $\omega_-=0$ happens if $\epsilon_\mathrm{\Delta}= \delta_\mathrm{s}/4$ that signals a possible instability in the system. However, this criterion needs to be dialed with a care since the spectrum $\omega_{\pm}$ has been derived using the plasma-pole approximation that has a limited applicability at low frequencies. Different instabilities of Dirac electron liquid ~\cite{MagneticInstability1,MagneticInstability2,Amperian1,Amperian2} enhanced by additional interactions mediated by spin-waves are outside the scope of this work.


The reduction of the classical dispersion Eqs.~\eqref{DispersionEq} to the Schrodinger-like ones, Eq.~\eqref{BdG1}, is an another important result of the paper and a key to the topological classification. The BdG dynamical matrix is non-Hermitian but paraunitary that is why its topological classification~\cite{TIClassification1,TIClassification2} differs compared to the one for Hermitian matrices~\cite{GeneralTopologicalClassification}. As we discuss in Appendix C, it belongs to D-class and is characterized by the integer (Chern) number. Here we follow a different root and argue that $K_\vec{q}$ and its truncated Hermitian version $H_\vec{q}$ (without coupling $Z_\vec{q}$ between positive and negative energy branches) are topologically equivalent.

It is instructive to introduce the modified BdG Hamiltonian $\bar{K}_\vec{q}[\alpha]$ by modifying $Z_\vec{q}\rightarrow \sin\alpha Z_\vec{q}$. At $\alpha=0$ it reduces to the truncated Hamiltonian $H_\vec{q}$ (supplemented by $-H_{-\vec{q}}^*$ that describes the spectrum of spurious negative energy branches). With increasing of $\alpha$ the modified BdG Hamiltonian evolves towards the full BdG one $K_q=\bar{K}_\vec{q}[\pi/2]$. The Chern numbers for each branch does not change during this evolution unless the spectrum experiences a band touching, $\bar{\omega}_-=0$ or $\bar{\omega}_-=\bar{\omega}_+$. Here $\bar{\omega}_\pm$ are two positive energy eigenstates of $\bar{K}_\vec{q}[\alpha]$ that are given by 
\begin{equation}
	\begin{split}
	\label{SpectrumModified}
	\bar{\omega}_{\pm}^2=\frac{\omega_\vec{q}^2+\epsilon_\vec{q}^2 + 2 |M_\vec{q}|^2 \cos^2\alpha}{2}\pm\\ \sqrt{ \frac{(\omega_\vec{q}^2-\epsilon_\vec{q}^2)^2}{4} + [4 \epsilon_\vec{q} \omega_\vec{q} +(\omega_\vec{q}-\epsilon_\vec{q})^2  \cos^2\alpha] |M_\vec{q}|^2}.
	\end{split} 
\end{equation}
The expression within the square root is obviously positive at any $\alpha$ yielding $\bar{\omega}_-\neq\bar{\omega}_+$. The discussed above stability condition $\epsilon_\mathrm{\Delta}<\delta_\mathrm{s}/4$ of the BdG Hamiltonian $K_\vec{q}$ ensures the absence of the band touching $\bar{\omega}_-=0$. We conclude that the spectra of $H_\vec{q}$ is smoothly connected with positive energy states of $K_\vec{q}$ that makes them topologically equivalent. 

As a result, the topology of spin-plasma waves can be addressed within the truncated two-band model, $H_\vec{q}$. It is intrinsically related to the momentum space texture for the unit vector $\vec{n}_\vec{q}=\vec{h}_\vec{q}/|\vec{h}_\vec{q}|$ defined within the Pauli matrix parametrization of the Hamiltonian $H_\vec{q}=\vec{h}_\vec{q}\cdot\hat{\mathbf{\sigma}} + h_0 \hat{1}$. The unit vector $\vec{n}_\vec{q}$ forms a topological skyrmion texture in momentum space, resulting in a band inversion for the dispersion curves for plasma and spin waves. It points down at $q=0$, lays in-plane around $q_*$ demonstrating the vortex-like texture, 
and flips up at $q\gg q_*$. Its topology is characterized by the Chern number that is defined as a momentum space integral over the Berry curvature $B_\vec{q}$,
\begin{equation}
	\label{Chern number}
	\mathcal{C}= \int \frac{d\vec{q}}{2\pi} B_\vec{q}, \;  \quad B_\vec{q}= \vec{n}_\vec{q} \cdot [\partial_{q_x} \vec{n}_\vec{q} \times \partial_{q_y} \vec{n}_\vec{q}].     
\end{equation} 
The Berry curvature characterizes the local geometry and its density in polar momentum coordinates $q B_q$ experience a maximum near $q_*$ as seen in Fig.~\ref{PlotResults}(c). The Chern numbers for two hybrid spin-plasma modes $\omega_\pm$ are equal to $\mathcal{C}_\pm=\pm1$. We discuss possible manifestations of nontrivial topology in Discussions, while their nonzero Berry curvature reflects itself in the thermal Hall effect. 

\section{Thermal Hall effect}
The nonzero Berry curvature of hybridized spin-plasma waves manifests in nonzero contribution to thermal Hall response. The contribution can be presented as follows~\cite{Matsumoto:PRL2011, Matsumoto:PRB2011}
\begin{equation}
	\label{ThermalConductivity}
	\kappa_\mathrm{xy}^\mathrm{B}=-\frac{T}{\hbar} \sum_{\vec{q},\nu=\pm 1}\left\{G[n_\mathrm{B}(\omega_{\nu \vec{q}})]  -\frac{\pi^2}{3} \right\}B_{\nu \vec{q}},
\end{equation}
where $B_{\nu\vec{q}}=\nu B_{\vec{q}}$ is the Berry curvature of hybrid spin-plasma waves with dispersion $\omega_{\nu\vec{q}}$. Here, $n_\mathrm{B}(\epsilon)$ is Bose-Einstein distribution function and 
$G(x)=(x+1)\ln^2[(1+x)/x]-\ln^2x-2\mathrm{Li}_{2}(-x)$ with Li$_{2}(x)$ as the polylogarithmic function. 

\begin{figure}[t]
	\vspace{-0.1in}
	\begin{center}
		\includegraphics[width=8.00 cm]{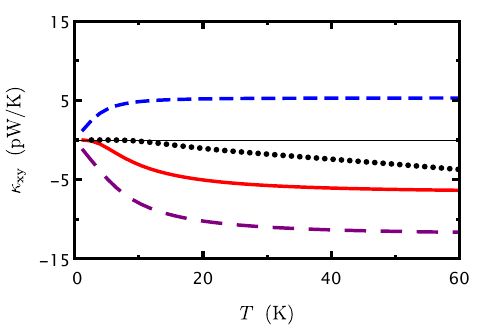}
		\caption{\label{thermal} Contribution of Dirac electrons (black dotes) and hybrid spin-plasma waves (solid red) to thermal Hall conductivity $\kappa_{xy}$. According to Eq.~(\ref{ThermalConductivity}), the latter is the sum of upper (short-dashed blue) and lower (long-range purple) hybrid modes.} 
	\end{center}
	\vspace{-0.1in}
\end{figure}

The temperature dependence of the contribution for the hybrid spin-plasma modes to the thermal Hall conductivity $\kappa_\mathrm{xy}^\mathrm{B}$ is presented in Fig.~(\ref{thermal}). The selective contributions of upper and lower modes have the opposite signs, but they do not compensate each other due to the thermal population imbalance between them. Their total contribution is considerably larger than the one for Dirac electrons that is evaluated in Appendix E and also presented in Fig.~(\ref{thermal}) for the comparison. Dirac electrons form the quantum degenerate liquid, $T\ll\epsilon_\mathrm{F}$, and their contribution is scaled by the small factor $T/\epsilon_\mathrm{F}$ that is not the case for the bosonic spin-plasma modes. An experimental observation of the nonlinear temperature dependence of $\kappa_\mathrm{xy}$ presented Fig.~(\ref{thermal}) will confirm the nonzero Berry curvature $B_\vec{q}$ for the hybridized spin-plasma waves.


\section{Discussions}
The calculated thermal Hall conductivity mediated by spin-plasma waves is $\kappa_{xy}\simeq 5\times 10^{-12}$ W$/$K. It exceeds the one that is predicted in systems with topological hybrid magnetoelastic waves (magnon-phonon modes) and is of order $\kappa_{xy}\simeq 10^{-13}$$\sim$$10^{-12}$ W$/$K~\cite{Zhang:PRL2019}. Both spin-plasma and spin-elastic waves represent hybridized and intertwined bosonic modes that makes the mechanism of the thermal Hall effect for them to be similar. In both cases the dominant contribution to the thermal conductivity comes from the vicinity of the avoided crossing where the Berry curvature is peaked. For the spin-plasma waves, the magnitude of the gap opened at the avoided crossings is comparable with the crossing energy. As a result, the only lower spin-plasma branch is well populated and the population imbalance favors the strong thermal Hall effect. For the case of magneto-elastic waves, the ratio between gap magnitude and the crossing energy is usually smaller that is why their contribution to the thermal Hall conductivity is also smaller.

The hallmark of the nontrivial topology is the presence of robust edge modes between regions with different Chern numbers. Flipping a direction of equilibrium magnetization in the magnet ($\vec{e}_\mathrm{z}\rightarrow-\vec{e}_\mathrm{z}$) reverses the precession of spin-waves ($a^-_{t\vec{r}}\rightarrow a^+_{t\vec{r}}$), inverts the phase winding factor in the matrix element ($M_\vec{q}\rightarrow M_\vec{q}^*$), and flips the topological Chern number ($\mathcal{C}_\pm\rightarrow -\mathcal{C}_\pm$). That is why a domain wall separating regions with opposite magnetizations is expected to host the protected edge spin-plasma modes. However, 
this prediction is based on the bulk-edge correspondence for the BdG dynamical matrix $K_\vec{q}$ and needs to be considered with a care. 

The Hamiltonian-like equations for spin waves have been derived within the plasma-pole approximation that is known to work very well only outside the continuum of single-particle excitations. As it seen in Fig.~2, the lower spin-plasma branch enters the continuum and acquires there the Landau damping. Coexistence of spin-plasma waves with the continuum is their essential and unavoidable feature since plasma waves are supported by Dirac electron liquid. However, if the avoided crossing is far away from the continuum, we expect the latter to have a little importance. Really, the edge modes represent a superposition of bulk ones mostly from the vicinity of the avoided crossing. The mixing with the overdamped modes from the continuum is minor and is not sufficient to break the bulk-boundary correspondence. It still can provide
a dissipation of edge spin-plasma modes that can be interpreted as \emph{edge Landau damping}. This regime,  $q_\mathrm{s}\ll q_*$, is the most favorable for observation of the edge spin-plasma waves and is achieved if $2d\ll\alpha$. 

The opposite limit with the avoiding crossing in the vicinity of the continuum is very delicate. At finite temperatures the continuum is  smoothed and the modes in the vicinity of the avoided crossing acquire the Landau damping. Their dissipative nature is essential and questions the range of validity of the bulk-boundary correspondence. However, the fate of the interplay between topology and dissipation is outside the scope of the present work and is postponed for future research. It should be noted that for the considered set of parameters $q/q_*\approx 0.4$ and the system is in the intermediate regime.



The considered here heterostructure can be realized in magnetically doped TIs~\cite{MagneticDoping1,MagneticDoping2,MagneticDoping3}, TI/ferromagnet interfaces,  e.g.  $\hbox{Tm}_3 \hbox{Fe}_5\hbox{O}_{12}$~\cite{MagneticInterface}, and magnetic topological insulators, e.g. $\hbox{MnBi}_2\hbox{Te}_4$~\cite{MagneticTI1,MagneticTI2,MagneticTI3,MagneticTI3,ProximityReview1,ProximityReview2}. Our results solely rely on the spin-momentum locking for Dirac electron liquid and do not require any fine-tuning. The material parameters determine the energy of avoided crossing and magnitude of the gap in its vicinity. The set of parameters is chosen for magnetic TI $\hbox{MnBi}_2\hbox{Te}_4$.  It support relatively uniform out-of-plane ordering of magnetic moments and their strong coupling with Dirac electrons. The latter is achieved due to the magnetic extension that implies the overlap of wave function for Dirac states and ordered magnetic moments. That is why magnetic TIs represent the most promising platform for observation of topological spin-waves.



\begin{acknowledgments} D. K. E acknowledges useful discussions with 
Olivier Bleu, Dimi Culcer and Oleg Sushkov. The research has been supported from the Australian Research Council Centre of Excellence in Future Low-Energy Electronics Technologies. M. K. acknowledges support from the Sharif University of Technology under Grant No. G960208 and Iran’s National Elite Federation.
\end{acknowledgments}

\bibliography{PMReferences}
\bibliographystyle{apsrev4-1_our_style}

\newpage
\begin{widetext}

\appendix

\section{Quantum field theory approach to the coupling between spin- and plasma waves}
In this appendix we present the derivation of the dispersion equation, Eq.~(\textcolor{red}{5}), using the quantum field theory formalism. The action of quantum Dirac liquid interacting with magnetic moments in the magnet can presented as sum of Fermionic $\mathcal{S}_\mathrm{F}$ and Bosonic $\mathcal{S}_\mathrm{B}$ actions supplemented by their coupling as $\mathcal{S}_\mathrm{FB}$ as follows   
\begin{equation}
\begin{split}
\mathcal{S}_{\mathrm{F}}& =\int d \tau d\vec{r} \; \bar{\psi}_{\tau \vec{r}}\left\{\partial_{\tau}+v [\hat{\vec{p}}\times \bm{\sigma}]_z + \Delta \sigma_z- \epsilon_{F}\right\}\psi_{\tau \vec{r}},  \\ 
\mathcal{S}_\mathrm{FB}& = \int d \tau d\vec{r}\; \bar{\psi}_{\tau \vec{r}}\left\{ i\phi_{\tau \vec{r}} +\Delta\bm{\sigma} \cdot \vec{l}_{\tau \vec{r}} \right\}\psi_{\tau \vec{r}}, \\
\mathcal{S}_{\mathrm{B}}& = \frac{\rho_\mathrm{s}}{2}\int d \tau d\vec{r} \left\{ [\partial_\tau \vec{l}_{\tau \vec{r}} \times \vec{l}_{\tau \vec{r}}]_z + \vec{l}_{\tau \vec{r}} \epsilon_{\hat{\vec{p}}} \vec{l}_{\tau \vec{r}}  \right\}    + \frac{1}{2}\int d\tau d\vec{r}d\vec{r}' \;V^{-1}_{\vec{r}-\vec{r}'} \phi_{\tau \vec{r}} \phi_{\tau \vec{r}'}.
\end{split}
\end{equation} 
Here $\psi_{\tau \vec{r}}=(\psi^{\uparrow}_{\tau \vec{r}},\psi^{\downarrow}_{\tau \vec{r}})^T$ is the spinor field describing Dirac electrons at the surface of a TI, and  $\tau$ is the imaginary (Matsubara) time. $\phi_{\tau \vec{r}}$ is the auxiliary bosonic field that has been introduced using the Hubbard-Stratonovich transformation to decouple repulsive Coulomb interactions as follows  
\begin{equation}
\int d\tau d\vec{r}d\vec{r}' V_{\vec{r}-\vec{r}'}\; \bar{\psi}_{\tau\vec{r}} \psi_{\tau\vec{r}} \bar{\psi}_{\tau\vec{r}'}\psi_{\tau\vec{r}'} =  \frac{1}{2}\int d\tau d\vec{r}d\vec{r}'\; V^{-1}_{\vec{r}-\vec{r}'} \phi_{\tau \vec{r}} \phi_{\tau \vec{r}'} + i \int d\tau d\vec{r}\; \phi_{\tau \vec{r}} \bar{\psi}_{\tau \vec{r}} \psi_{\tau \vec{r}}.      
\end{equation} 
Its physical meaning is the scalar potential and the imaginary unit $i$ in front of its coupling with Dirac liquid is a mathematical peculiarity of the imaginary time formalism. Really, the Wick rotation transform the covariant derivative $\partial_t+\phi$ to $\partial_{\tau}+i \phi$, and the corresponding unit $i$ emerges. It is instructive to reorganize magnetisation vector $\vec{l}_{\tau \vec{r}}$ in complex fields $l^-_{\tau \vec{r}}=l_{\tau \vec{r}}^\mathrm{x}-i l_{\tau \vec{r}}^\mathrm{y}$ and $l^+_{\tau \vec{r}}=l_{\tau \vec{r}}^\mathrm{x}+i l_{\tau \vec{r}}^\mathrm{y}$ and group all bosonic fields into $f_{\tau \vec{r}}=\{i e \phi_{\tau \vec{r}}, \Delta l^-_{\tau \vec{r}}, \Delta l^+_{\tau \vec{r}} \}$. By Fourier transformation the bosonic action can be presented as follows
\begin{equation}
\mathcal{S}_\mathrm{B}=\frac{1}{2} \sum_q f_q^\dagger(-L^0_q) f_q, \quad \quad  L^0_q=\mathrm{diag}\left[V_{\vec{q}}^{-1}, \frac{\rho_\mathrm{s}}{\Delta^2}(i p_n -\epsilon_\vec{q}), \frac{\rho_\mathrm{s}}{\Delta^2} (-i p_n -\epsilon_\vec{q}) \right].    
\end{equation}
Here $q=\{i p_n, \vec{q}\}$ includes both momentum $\vec{q}$ and Bosonic Matsubara frequency $p_n=2\pi n T$. Importantly, Bosonic fields $f_q$ do not interact directly with each other, but only with Dirac liquid as follows
\begin{equation}
\mathcal{S}_\mathrm{FB}=\int d\tau d\vec{r}\; m_{\tau \vec{r}} \cdot f_{\tau \vec{r}} = \sum_q\; m_{-q} \cdot f_q, \quad\quad m_{\tau \vec{r}}=\left\{\rho_{\tau \vec{r}}, s^+_{\tau \vec{r}}, s^-_{\tau \vec{r}} \right\}.       
\end{equation}
Here we have introduced the vector $m_{\tau \vec{r}}$ composed of the matter fields in the similar manner as it is done in the paper. The action is quadratic in respect to Fermionic fields and they can be integrating out. Expanding the resulting action up to the second order in Bosonic fields $f_{\tau \vec{r}}$ around the trivial saddle point $\bar{f}_q=0$ we get
\begin{equation}
\label{IntegrationOut}
\mathcal{S}'_\mathrm{B}=\frac{1}{2} \sum_q f_q^\dagger(-L^0_q+\hat{\Pi}_q) f_q.    
\end{equation}
Here $\hat{\Pi}_q$ is the generalized response functions between matter fields. Having established the effective description of bosonic fields $\mathcal{S}'_\mathrm{B}$, the saddle point of the quantum action, Eq.~\eqref{IntegrationOut}, corresponds to the classical equations of motion that are given by $(L^0_q-\Pi_q)f_q=0$. After the analytical continuation, $ip_n\rightarrow \omega+i \delta$ and $i\phi_{i p_n\vec{q}} \rightarrow \phi_{\omega\vec{q}}$, the resulting equation matches with Eq.~(\ref{DispersionEq}) that has been derived in the main text within the classical picture for the coupling between spin- and plasma- waves.

\section{The linearized Landau-Lifshitz-Gilbert equation}
In this appendix we present a detailed derivation of the dispersion equation, Eq.~(\textcolor{red}{5}), for hybrid spin-plasma waves. The dynamics of magnetic moments directed along $\vec{n}_{t\vec{r}}$ follow the Landau-Lifshitz-Gilbert equation~\cite{LL} given by
\begin{equation}
	\label{LLSM1} 
	\rho_\mathrm{s} \partial_t \vec{n}_{t\vec{r}} = \left[B_{t\vec{r}} \times n_{t\vec{r}}\right], \quad \quad H_\vec{n}=\int d\vec{r}\left\{\frac{\rho_\mathrm{s}}{2} \left[\frac{|\nabla \vec{n}_{t\vec{r}}|^2}{2 m_\mathrm{s}} + \delta_\mathrm{s} \left((n_{t\vec{r}}^x)^2+ (n_{t\vec{r}}^y)^2\right) \right] + \Delta \vec{n}_{t\vec{r}} \vec{s}_{t\vec{r}}  \right\}
\end{equation}
Here $B_{t\vec{r}}=-\delta H_\vec{n}/\delta n_{t\vec{r}}$ is usually interpreted as effective magnetic field that induces the precession of magnetic moments. The first term in the magnetic energy $H_\vec{n}$ is intrinsic for the magnet with $\rho_\mathrm{s}$ is the density of magnetic moments and $m_\mathrm{s}$ parametrizes their gradient energy. It is assumed that the magnet has the easy-axis anisotropy and energy $\delta_\mathrm{s}$ determines its strength. The second term in $H_\vec{n}$ describes the interaction between magnetic moments and spin density  $\vec{s}_{t\vec{r}}=s^\mathrm{x}_{t\vec{r}} \vec{e}_\mathrm{x}+ s^\mathrm{y}_{t\vec{r}} \vec{e}_\mathrm{y}$ of Dirac liquid at the surface of topological insulator (TI). The energy $\Delta$ determines their coupling strength.

The anisotropy favors magnetic moments to be ordered perpendicular to the TI surface, e.g. in the $\vec{e}_\mathrm{z}$-direction. As a result, propagating small-amplitude magnetic fluctuations have only in-plane component $\vec{l}_{t\vec{r}}=l^\mathrm{x}_{t\vec{r}} \vec{e}_\mathrm{x}+ l^\mathrm{y}_{t\vec{r}} \vec{e}_\mathrm{y}$ and are known as spin-waves or magnons. The linearization  $\vec{n}_{t\vec{r}}=\vec{e}_\mathrm{z}+\vec{l}_{t\vec{r}}$ of the Landau-Lifshitz-Gilbert equation, Eq.~(\ref{LLSM1}), results in the Eq.~(\ref{LL1}) from the main paper that is given by
\begin{equation}
	\label{LLSM2} 
	\rho_\mathrm{s} \left[\partial_t \vec{l}_{t\vec{r}} \times \vec{e}_\mathrm{z}\right]= \rho_\mathrm{s}\epsilon_{\hat{\vec{p}}} \vec{l}_{t \vec{r}} + \Delta \vec{s}_{t\vec{r}}
\end{equation}
Here $\epsilon_\vec{p}=\delta_\mathrm{s}+\vec{p}^2/2 m_s$ is the dispersion of spin-waves.  They are coupled with the spin density $\vec{s}_{t\vec{r}}$ of Dirac liquid and can be excited by its oscillations.

\section{The response functions of the helical Dirac electron liquid}
This appendix presents derivation of the response functions of Dirac electrons at the surface of a topological insulator. Dirac electrons can be described by the following Hamiltonian 
\begin{equation}
H=v [\vec{p}\times \bm{\sigma}]_z + \Delta \sigma_z- \epsilon_{F}. 
\end{equation}
Here $v$ and $\epsilon_\mathrm{F}$ are velocity and Fermi energy of Dirac electrons. $2 \Delta$ is the gap between Dirac valence ($\gamma=-1$) and conduction ($\gamma=1$) bands $\epsilon_{\gamma \vec{p}}=\gamma\epsilon_\vec{p}$ with $\epsilon_\vec{p}=\sqrt{v^2 p^2+\Delta^2}$ that is induced by coupling to the equilibrium static out-of-plane magnetization. Their spinor wave functions are given by 
\begin{equation}
|+,\vec{p}\rangle =\begin{pmatrix} \cos(\frac{\theta}{2}) \\ i \sin(\frac{\theta}{2}) e^{i \phi_\vec{p}}\end{pmatrix}, \quad \quad |-,\vec{p}\rangle =\begin{pmatrix} \sin(\frac{\theta}{2}) \\ - i \cos(\frac{\theta}{2}) e^{i \phi_\vec{p}}\end{pmatrix}.
\end{equation}
Here $\phi_\vec{p}$ is the polar angle for vector $\vec{p}$ and $\cos(\theta)=\Delta/\epsilon_\vec{p}$. 

The powerful approach for analytical calculation of the polarization operator $\hat{\Pi}_{00}(\omega,\vec{q})$ has been developed in Refs.~\cite{GraphenePlasmons1,GraphenePlasmons2,GraphenePlasmonsReview} and can be extended to other response functions $\hat{\Pi}(\omega,q)$. However, in the present Letter we are interested only in the long-wave $q\ll p_\mathrm{F}$ and low-frequency $\omega\ll \epsilon_\mathrm{F}$ limit. In this regime only electron-hole excitations in the vicinity of the Fermi level for Dirac particles are essential and calculations can be drastically simplified.

At first, transitions between Dirac valence and conduction bands for surface states can be neglected. Without any loss of generality, we assume that the Fermi level of Dirac electrons is in the Dirac conduction band $\epsilon_\mathrm{F}>0$.  As a result, the generalized response functions $\Pi^{\alpha\beta}_{\omega \vec{q}}$ with $\alpha, \beta \in \{0, +, -\}$ is given by  
\begin{equation}
\label{Pi}
\Pi^{\alpha \beta}_{\omega \vec{q}}=\sum_{\vec{p}} \langle +,\vec{p}_-|\sigma_\alpha |+, \vec{p}_+\rangle \langle +,\vec{p}_+|\sigma_\beta |+, \vec{p}_{-}\rangle \frac{n_\mathrm{F}(\epsilon_{\vec{p}_-})-n_\mathrm{F}(\epsilon_{\vec{p}_+})}{\omega+\epsilon_{\vec{p}_-}-\epsilon_{\vec{p}_+}+i \delta}\equiv \sum_{\vec{p}} \Lambda_{\vec{p}\vec{q}}^{\alpha \beta}\frac{n_{\vec{p}\vec{q}}}{\omega-\epsilon_{\vec{p}\vec{q}}+i \delta}.  
\end{equation}
Here $\vec{p}_\pm =\vec{p}\pm \vec{q}/2$ and $n_\mathrm{F}(\epsilon_\vec{p})$ is the Fermi-Dirac distribution function at $T=0$. It is equal to $n_\mathrm{F}(\epsilon_\vec{p})=1$ within the Fermi sea $p<p_\mathrm{F}$ and $n_\mathrm{F}(\epsilon_\vec{p})=0$ outside it $p>p_\mathrm{F}$. The explicit form of the matrix elements product $\Lambda_{\vec{p}\vec{q}}^{\alpha \beta}$ is given by
\begin{equation}
\begin{split}
\Lambda_{\vec{p}\vec{q}}^{00}=\frac{1}{2}\left(1+\frac{\Delta^2+v^2 \vec{p}_- \vec{p}_+}{\epsilon_{\vec{p}_-} \epsilon_{\vec{p}_+}} \right), \quad \quad \quad \quad \quad \quad \quad \quad \quad \quad \quad \quad  \Lambda_{\vec{p}\vec{q}}^{-+}=\frac{1}{4}\left(1+\frac{\Delta}{\epsilon_{\vec{p}_+}}-\frac{\Delta}{\epsilon_{\vec{p}_-}}-\frac{\Delta^2}{\epsilon_{\vec{p}_+}\epsilon_{\vec{p}_-}} \right), \\
\Lambda_{\vec{p}\vec{q}}^{0\pm}= \pm \frac{i}{4}\left( \frac{v p_- e^{\pm i \phi_{\vec{p}_-}} }{\epsilon_{\vec{p}_-}} + \frac{v p_+ e^{\pm i \phi_{\vec{p}_+}}}{\epsilon_{\vec{p}_+}} \mp \frac{\Delta v q e^{\pm i \phi_\vec{q}}}{\epsilon_{\vec{p}_-} \epsilon_{\vec{p}_+} }  \right), \quad \quad \quad \quad \quad  \quad  
\Lambda_{\vec{p}\vec{q}}^{\pm\pm}= \frac{1}{4} \frac{v^2 p_- p_+ e^{\pm i (\phi_{\vec{p}_-}+ \phi_{\vec{p}_+})}}{\epsilon_{\vec{p}_-} \epsilon_{\vec{p}_+}}.	\end{split}
\end{equation} 
Importantly, two of them, $\Lambda_{\vec{p}\vec{q}}^{0\pm}$ and $\Lambda_{\vec{p}\vec{q}}^{\pm \pm}$, have the phase winding factor $e^{i \phi_\vec{q}}$. Its presence is clearly seen if we rewrite them as follows
\begin{equation}
\begin{split}
\Lambda_{\vec{p}\vec{q}}^{0\pm}= \pm \frac{i e^{\pm i \phi_\vec{q}}}{4}\left( \frac{v (p e^{\pm i \phi}-\frac{q}{2}) }{\epsilon_{\vec{p}_-}} + \frac{v ( p e^{\pm i \phi}+\frac{q}{2}) }{\epsilon_{\vec{p}_+}} \mp \frac{\Delta v q}{\epsilon_{\vec{p}_-} \epsilon_{\vec{p}_+}}\right), \quad \quad \Lambda_{\vec{p}\vec{q}}^{\pm\pm}= \frac{e^{\pm 2 i\phi_\vec{q}}}{4} \frac{v^2 (p e^{\pm i \phi}+\frac{q}{2}) (p e^{\pm i \phi}-\frac{q}{2})}{\epsilon_{\vec{p}_-} \epsilon_{\vec{p}_+}}.
\end{split}
\end{equation}
Here $\phi$ is the angle between momenta $\vec{q}$ and $\vec{p}$. After the shift $\phi_\vec{p}=\phi_\vec{q}+\phi$ the integration measure in Eq.~(\ref{Pi}) transforms as $pdp d\phi_\vec{p}\rightarrow pdp d\phi$. As a result, the phase winding factor $e^{i \phi_\vec{q}}$ can be taken out of the integral and becomes essential ingredient of $\Pi_{\vec{p}\vec{q}}^{0\pm}$ and $\Pi_{\vec{p}\vec{q}}^{\pm \pm}$.

The condition $q\ll p_\mathrm{F}$ allows to make the further simplifications
\begin{equation}
\epsilon_{\vec{p}\vec{q}}=u q \cos(\phi), \quad\quad n_{\vec{p} \vec{q}} = q \cos(\phi) \delta (p-p_\mathrm{F}). 
\end{equation}
Here $u=v h$ is the Fermi velocity of massive Dirac electrons and $h=v p_\mathrm{F}/\epsilon_\mathrm{F}$. Its physical meaning is the in-pane component of spin for Dirac electrons. If we approximate the product of matrix elements $\Lambda^{\alpha \beta}_{\vec{p}\vec{q}}$ by its value at $q=0$ and $p=p_\mathrm{F}$ we get 
\begin{equation}
\Lambda^{00}_{\vec{p}\vec{q}}=1, \quad \quad \Lambda^{-+}_{\vec{p}\vec{q}}=\frac{h^2}{4}, \quad \quad \Lambda^{0\pm}=\pm \frac{i h e^{i\phi_\vec{q}}}{2} (\cos(\phi)+i \sin(\phi)), \quad \quad \Lambda^{\pm\pm}_{\vec{p}\vec{q}}=\frac{h^2 e^{\pm 2 i\phi_\vec{q}}}{4} (\cos(2 \phi)+ i \sin(2 \phi)).           
\end{equation}
The odd terms in $\phi$ (the ones proportional to $\sin(\phi)$ or $\sin(2\phi)$) vanish after the angle integration and can be omitted. As a result, the spin-charge polarization functions can be presented as 
\begin{equation}
\Pi^{00}_{\omega \vec{q}} =N_\mathrm{F} I_1(\Omega), \quad \quad \Pi^{-+}_{\omega \vec{q}}=\frac{h^2}{4} N_\mathrm{F} I_1(\Omega), \quad \quad \Pi^{0\pm}_{\omega \vec{q}}=\pm \frac{i h e^{i\phi_\vec{q}}}{2} N_\mathrm{F} I_2(\Omega), \quad \quad \Pi^{\pm\pm}_{\omega \vec{q}}=\frac{h^2 e^{\pm 2 i\phi_\vec{q}}}{4} N_\mathrm{F} I'(\Omega).           
\end{equation}
Here $\Omega=\omega/uq$ and $N_\mathrm{F}=\epsilon_\mathrm{F}/2\pi \hbar^2 u$ is the density of states at the Fermi level. The functions $I_n(\Omega)$ and $I'(\Omega)$ are defined as
\begin{equation}
I_n(\Omega)=\int \frac{d \phi}{2\pi} \frac{\cos^n \phi}{\Omega-\cos \phi+i \delta}, \quad \quad \quad I'(\omega)=2 I_3(\omega)-I_1(\omega).
\end{equation}
They can be evaluated with the help of recurrence relations as follows 
\begin{equation}
I_{n+1}(\Omega)=-\frac{\cos^2(\pi n/2) n!}{2^n (n/2)!} + \Omega I_n(\Omega), \quad\quad \quad I_0(\Omega)= \Theta(|\Omega|-1)\frac{\mathrm{sgn}[\Omega]}{\sqrt{\Omega^2-1}}-i \Theta(|\Omega|-1) \frac{1}{\sqrt{1-\Omega^2}}. 
\end{equation}
Importantly, the relations $I_2(\Omega)=\Omega I_1(\Omega)$ and $h \Omega= \omega/vq $ ensure the connection between density-density and spin-density susceptibilities
\begin{equation}
\Pi_{\omega \vec{q}}^{\pm0}=\pm\frac{i e^{\pm i\phi_\vec{q}}}{2}\frac{\omega}{vq} \Pi_{\omega \vec{q}}^{00},    
\end{equation}
that is the cornerstone of our theory and ensures the nontrivial topology of the hybrid spin-plasma modes. 

\section{Topological classification of Bogoliubov-de Gennes (BdG) Hamiltonian}
In this appendix we briefly overview the spectrum of the BdG dynamical matrix $\hat{K}_\vec{q}$ and its topological classification. The explicit form of $\hat{K}_\vec{q}$ is given by
\begin{equation}
\hat{K}_\vec{q}=\begin{pmatrix} \hat{H}_\vec{q} & \hat{Z}_\vec{q} \\ -\hat{Z}^\dagger_\vec{q} & -\hat{H}^*_\vec{q} \end{pmatrix}, \quad \quad \quad \hat{H}^\mathrm{BdG}_\vec{q}=\begin{pmatrix} \hat{H}_\vec{q} & \hat{Z}_\vec{q} \\ \hat{Z}^\dagger_\vec{q} & \hat{H}^*_\vec{q} \end{pmatrix}.
\end{equation}
Here we have also introduced the BdG Hamiltonian $\hat{H}_\vec{q}^\mathrm{BdG}=\hat{\Sigma}_\mathrm{z} \hat{K}_\vec{q}$ with $\Sigma_\mathrm{z}$ is one of the generalized Pauli matrices
\begin{equation}
\Sigma_x=\begin{pmatrix} 0 & \hat{1} \\ \hat{1} & 0 \end{pmatrix}, \quad \quad \quad \Sigma_y=\begin{pmatrix} 0 & -i \; \hat{1} \\ i \; \hat{1} & 0 \end{pmatrix}, \quad \quad \quad  \Sigma_z=\begin{pmatrix} 1 & 0 \\ 0 & -\hat{1} \end{pmatrix}.
\end{equation}
It should be noted that $H_\vec{q}^\mathrm{BdG}$ really plays the role of the Hamiltonian if we quantize spin- and plasma- waves. The topological classification is based on the symmetries of $\hat{K}_\vec{q}$ and $\hat{H}_\vec{q}^\mathrm{BdG}$ as at has been recently discussed~\cite{TIClassification1,TIClassification2}. Both of them enjoy \emph{only} the particle-hole symmetry as $\mathcal{C}\hat{K}_\vec{q}\mathcal{C}^{-1}=-\hat{K}_{-\vec{q}}$ and $\mathcal{C}\hat{H}_\vec{q}^\mathrm{BdG}\mathcal{C}^{-1}=\hat{H}_{-\vec{q}}^\mathrm{BdG}$.  Here  $\mathcal{C}=\Sigma_\mathrm{x} \mathcal{K}$ where $\mathcal{K}$ is the complex conjugation operator. 
Since $\mathcal{C}^2=1$, the dynamical matrix $K_\vec{q}$ belongs to class D, while the BdG Hamiltonian $\hat{H}_\vec{q}^\mathrm{BdG}$ falls in class CI. Both of them  ensures the same classification of the spectrum in terms of the integer topological Chern number $\mathbb{Z}$. In the main text of the paper we argue that the spectra of $K_\vec{q}$ and its truncated hermitian Hamitonian $H_\vec{q}$ are smoothly connected that ensures them to be topologically equivalent. The Hamiltonian $\hat{H}_\vec{q}$ represents a bosonic analogue of quantum Hall effect (or Haldane model with no non-spatial symmetry) and therefore belongs to the same topological class characterized by integer Chern numbers $\mathbb{Z}$.

The dynamical matrix $K_\vec{q}$ is non-Hermitain that modifies the Chern number calculation differs compared to the one for Hermitian matrices~\cite{GeneralTopologicalClassification}. It is instructive to discuss it in more detail. Due to the particle-hole symmetry, solutions of the eigenvalue problem $K_\vec{q} \psi_{\omega \vec{q}} = \omega \psi_{\omega \vec{q}}$ appear in pairs $|+,\nu,\vec{q} \rangle$ and $|-,\nu,\vec{q}\rangle$ and are not independent from each other. The states $|+,\nu, \vec{q},\rangle$ have positive energies $\omega_{+,\nu}=\omega_{\nu}$ and are labeled are by $\nu=\pm 1$. The states $|-,\nu, \vec{q}\rangle$  have inverted energies $\omega_{-\nu}=-\omega_{\nu}$ and their wave functions are connected by the particle-hole transformation $|-,\nu, \vec{q}\rangle=\mathcal{C}  |+,\nu,\vec{q} \rangle$. The dynamical matrix $K_\vec{q}$ is not Hermitian, but the BdG Hamiltonian $\hat{H}_\vec{q}^\mathrm{BdG}=\Sigma_\mathrm{z} \hat{K}_\vec{q}$ Hermitian. That is why adjoint states are defined as $\overline{\langle \pm, \nu, \vec{q}|}=\langle \pm, \nu, \vec{q}|\Sigma_z$ and are normalized as $\overline{\langle \pm, \nu, \vec{q}}| \Sigma_z| \pm, \nu, \vec{q}\rangle=\pm 1$. The adjoint state is also involved in the definition of Berry connection and curvature as well as the Chern number $C_{\pm\nu}$ as follows
\begin{equation}
\label{BdGGeometry}
A_{\pm\nu \vec{q}}=\overline{\langle \pm \nu \vec{q}|} i \nabla_\vec{q}|\pm \nu \vec{q}\rangle, \quad \quad B_{\pm\vec{q}}=[\nabla_\vec{q}\times A_{\pm \gamma\vec{q}}]_\mathrm{z}  \quad \quad C_{\pm\nu} = \int \frac{d\vec{q}}{2\pi} B_{\pm\nu\vec{q}}. 
\end{equation}
Importantly, the Berry curvature for positive and negative branches is the same $A_{-\gamma, \vec{q}}=\langle -\nu \vec{q}| \Sigma_\mathrm{z} i \nabla_\vec{q}|-\nu \vec{q}\rangle=\langle \nu \vec{q}|\; \Sigma_\mathrm{x} C \; \Sigma_z i \nabla_\vec{q} \; \sigma_\mathrm{x} C \;|\nu\vec{q}\rangle=A_{\nu\vec{q}}$. This ensures that topological Chern numbers for the negative and positive energy states do match each other $C_{-\nu}=C_{\nu}$. As a result, only the latter can be considered as we do in the paper. The expressions (\ref{BdGGeometry}) have been used to calculate the Berry curvature $B_\vec{q}$ for BdG Hamiltonian $K_\vec{q}$ that is presented in Fig.~2 of the paper. 

\section{Thermal Hall effect}

This Appendix presents the results of the thermal Hall conductivity $\kappa_\mathrm{xy}$ of the interface between topological insulator and a magnet. It corresponds to the linear response relation  $\vec{J}_{x}^{Q}=- \kappa_\mathrm{xy} \nabla_\mathrm{y} T$ between the heat current $\vec{J}_{x}^{Q}$ and temperature gradient $\nabla_\mathrm{y} T$. Upon the quantization of the dynamics of hybrid spin- and plasma-waves, they become bosonic modes (that are usually referred as magnons and plasmons). In our model there are two contributions to thermal Hall conductivity $\kappa_{xy}=\kappa^\mathrm{F}_{\mathrm{xy}}+\kappa^\mathrm{B}_{\mathrm{xy}}$, the fermionic Dirac electrons $\kappa^\mathrm{F}_{\mathrm{xy}}$ and bosonic modes~$\kappa^\mathrm{B}_{\mathrm{xy}}$.

The contribution of Dirac states is given by~\cite{Vafek:PRB2001} 
\begin{equation}
	\kappa^\mathrm{F}_{\mathrm{xy}}=-\frac{\hbar}{e^2 T}\int d\epsilon ~(\epsilon-\epsilon_\mathrm{F})^2\sigma_{xy}(\epsilon) n_\mathrm{F}'(\epsilon-\epsilon_\mathrm{F}),
\end{equation}
where $n_\mathrm{F}(\epsilon)$ is the Fermi-Dirac distribution function, and $\sigma_{xy}(\epsilon)$ is the anomalous Hall conductivity at energy $\epsilon$:
\begin{equation}
	\sigma_\mathrm{xy}(\epsilon)=\frac{e^2}{\hbar}\sum_{\vec{p},\gamma=\pm1} \Omega^\mathrm{D}_{\gamma}(\vec{p})~\Theta(\epsilon-E_{\gamma}(\vec{p})),
\end{equation}
with $\Theta(x)$ as the Heaviside function. Here, $\Omega^\mathrm{D}_{\gamma}(\vec{k})$ is the Berry curvature of valence ($\gamma=-1$) and conduction ($\gamma=1$) bands of Dirac states. The temperature dependence of $\kappa^\mathrm{F}_{\mathrm{xy}}$ calculated for the set of parameters presented in the Letter is shown in Fig.~(\ref{thermal}). The contribution of degenerate quantum electron liquid is linear and even at high temperatures ($T\approx 20\;\hbox{K}$) it yields values of order of $10^{-12}$ W$/$K. 

The nonzero Berry curvature of hybridized spin-plasma waves manifests in nonzero contribution to thermal Hall response. The contribution can be presented as follows~\cite{Matsumoto:PRL2011, Matsumoto:PRB2011}
\begin{equation}
	\label{ThermalConductivityAp}
	\kappa_\mathrm{xy}^\mathrm{B}=-\frac{T}{\hbar} \sum_{\vec{q},\nu=\pm 1}\left\{G[n_\mathrm{B}(\omega_{\nu \vec{q}})]  -\frac{\pi^2}{3} \right\}B_{\nu \vec{q}},
\end{equation}
where $B_{\nu\vec{q}}=\nu B_{\vec{q}}$ is the Berry curvature of hybrid spin-plasma waves with dispersion $\omega_{\nu\vec{q}}$. Here, $n_\mathrm{B}(\epsilon)$ is Bose-Einstein distribution function and 
$G(x)=(x+1)\ln^2[(1+x)/x]-\ln^2x-2\mathrm{Li}_{2}(-x)$ with Li$_{2}(x)$ as the polylogarithmic function. 

The temperature dependence of $\kappa_\mathrm{xy}^\mathrm{B}$ is presented in Fig.~(\ref{thermal}). The flatness of the dispersion for hybrid spin-plasma waves results in their strong impact $\kappa_\mathrm{xy}^\mathrm{B}$ in the thermal Hall conductance. It is about $5\times 10^{-12}$ W$/$K, which is larger than its electronic part $\kappa_\mathrm{xy}^\mathrm{F}$. 
An experimental observation of the nonlinear temperature dependence of $\kappa_\mathrm{xy}$ presented Fig.~(\ref{thermal}) will confirm the nontrivial topology of the hybridized spin-plasma waves.  

It also should be mentioned that the calculated thermal Hall conductivity mediated by spin-plasma waves is more that one order of magnitude larger than one that is predicted in systems with topological hybrid spin- and elastic- waves (magnon-phonon modes) and is of order $10^{-13}$ W$/$K~\cite{Zhang:PRL2019}.

\end{widetext}

\end{document}